\documentclass[]{rAMF2e}

\begin{document}

%\articletype{GUIDE}

{\Large {\bf New analytic approach to address Put - Call parity violation due to discrete dividends}}

\vspace{4mm}

ALEXANDER BURYAK and IVAN GUO

\vspace{4mm}

aburyak@bigpond.net.au

%\received{v1.0 released 15 November 2010}

% \maketitle

\begin{abstract}
The issue of developing simple Black-Scholes type approximations for pricing European options with large discrete dividends was popular since early 2000's with a few different approaches reported during the last 10 years. Moreover, it has been claimed that at least some of the resulting expressions represent high-quality approximations which closely match results obtained by the use of numerics.

In this paper we review, on the one hand, these previously suggested Black-Scholes type approximations and, on the other hand, different versions of the corresponding Crank-Nicolson numerical schemes with a primary focus on their boundary condition variations. Unexpectedly we often observe substantial deviations between the analytical and numerical results which may be especially pronounced for European Puts. Moreover, our analysis demonstrates that any Black-Scholes type approximation which adjusts Put parameters identically to Call parameters has an inherent problem of failing to detect a little known Put-Call Parity violation phenomenon. To address this issue we derive a new analytic approximation which is in a better agreement with the corresponding numerical results in comparison with any of the previously known analytic approaches for European Calls and Puts with large discrete dividends.


\end{abstract}

\section{Introduction}

The issue of accurately pricing European options with large discrete dividends was popular in early 2000's: \cite{BenederW2001,Frishling2002,BosV2002,BosGS2003,HaugHL2003,VellekoopN2006} and is currently attracting a renewed attention [see, e.g., \cite{AmaroDF2009,SivenSP2009,VeigaW2009}] which at least in part may be attributed to the recent large corrections of stock markets and the related increases in projected dividend yields due to declining share prices.

It is interesting to note that in \cite{BenederW2001,BosV2002,BosGS2003} simple structural modifications of strike, spot or volatility parameters (based on forecast dividends and actual values of other parameter) in the conventional Black-Scholes (BS) formula (see e.g. p. 259 of \cite{Hull2006}) were suggested. Moreover, it has been claimed that some of the resulting expressions represent high-quality approximations which closely match the corresponding numerical results based on Crank-Nicolson algorithm [e.g. in \cite{BosV2002,BosGS2003}]. Later the approximations of \cite{BenederW2001,BosV2002,BosGS2003} were criticised by \cite{HaugHL2003,VellekoopN2006}, where more accurate, but essentially numerical approaches were suggested. These numerical approaches will not be the focus of our investigation.

We start our analysis by, on the one hand, reviewing the previously reported analytical approximations and, on the other hand, choosing between different versions of Crank-Nicolson numerical schemes with a primary focus on their boundary condition variations. Then, after clarifying our choice of the benchmark numerical scheme, we make an initial qualitative comparison of the existing analytical approximations for European options with large discrete dividends with Crank-Nicolson based modelling results. Unexpectedly we observe substantial deviations between the analytical and numerical results for both European Call and Put options with the discrepancy for Puts being more qualitatively significant. This can be explained by a little known Put-Call Parity violation phenomenon, which, according to our knowledge, was first mentioned by \cite{HaugHL2003} in the literature. This Parity violation is only present in discrete (non-continuous) dividend models and is due to dividend-induced changes in Put option payout shapes and resulting deviations from the conventional log-normal process induced distributions. In the simplest case of a single (large) dividend payout, one can readily obtain an exact analytic expression for the calculation of Parity violation values. However, no analytic results have been reported for a multi-dividend case.  We address this issue by developing a new higher quality analytic approximation for both Calls and Puts which, among other features, is capable of taking into account the parity violation adjustment. After outlining the details of the new method, we perform a final quantitative comparison of existing and newly developed analytic methods with CN numerics and confirm that our new algorithm indeed outperforms other analytic approaches.

\section{Methods}\label{Methods}

In this paper we concentrate on analysing the stock process $S_t$ which jumps down by the amounts of dividend $d_i$ at the respective times $t_i$, and follows a geometric Brownian motion with flat volatility $\sigma$ at other times (i.e. before, in between and after dividend payouts):
\begin{equation} \label{stockprocess}
dS_t=\left(rS_t-\sum_{0<t_i\leq T} d_i \delta(t-t_i)\right)dt + \sigma S_t dW_t
\end{equation}
where $r$ is the risk-free interest rate, $\delta$ is the Dirac delta function and $W_t$ is a Wiener process (see e.g. \cite{Hull2006} for details).

We start this section by reviewing existing analytic approximations for options with discrete dividends. Then we move to discussing finer details of the utilised benchmark numerical method and compare the corresponding numerical results with different analytic approximations. Finally we discuss the observed discrepancies and outline a new solution which successfully addresses them.

\subsection{Review of existing of Black-Scholes-type analytic approximations}

In early 2000's the work of \cite{Frishling2002} initiated the discussion by highlighting the issue of significant differences between conventional analytical BS results for European options with discrete dividends and the corresponding numerical results. It points out that put/call options for stocks with discrete dividends do not allow the direct use of the conventional BS formulas:
\begin{equation} \label{BS_conven}
\begin{array}{l}
{\displaystyle
{C=S_0 \Phi(b_1)-K \exp{(-r T)} \Phi(b_2),}}
\\
\\
{\displaystyle
{P=K \exp{(-rT)} \Phi(b_2) - S_0 \Phi(-b_1),}}
\end{array}
\end{equation}
where we use the usual notations, $C$  and  $P$ are a Call and a Put, respectively; $S_0$ is the corresponding current stock price (spot price), $K$ is the strike, $T$ is the term (time to maturity) of the option, $\Phi$ is the cumulative Gaussian distribution function, and $b_1$ and $b_2$ are also given by their conventional expressions: $b_1 = [\ln{(S_0/K)} + (r+\sigma^2/2) T]/(\sigma \sqrt{T})$ and $b_2 = b_1 - \sigma \sqrt{T}$.

Expressions (\ref{BS_conven}) can be easily extended to allow consideration of stocks with continuous dividend yields by making the change  $S_0 \rightarrow S_0 \exp{(-q T)}$. This, in fact, defines arguably the most conventional (although not the most accurate) approximation for calculation of options for stocks with discrete dividends, corresponding to the underlying process given by Eq. (\ref{stockprocess}). Below we will refer to it as the vanilla spot adjustment approximation or simply \emph{spot approximation}. Instead of calculating a dividend yield $q$ and applying the $S_0 \rightarrow S_0 \exp{(-q T)}$ change one can equivalently subtract the present value of dividends $\displaystyle {D \equiv \sum_{0<t_i\le T}{d_i \exp{(-r t_i)}}}$, from $S_0$ to get $\displaystyle {\tilde{S}_0 = S_0 - \sum_{0<t_i\le T}{d_i \exp{(-r t_i)}}}$ and change $S_0$ into $\tilde{S}_0$ everywhere in Eq. (\ref{BS_conven}):
\begin{equation} \label{BS_spot}
\begin{array}{l}
{\displaystyle
{C=\tilde{S}_0 \Phi(b_1)-K \exp{(-r T)} \Phi(b_2),}}
\\
\\
{\displaystyle
{P=K \exp{(-rT)} \Phi(b_2) - \tilde{S}_0 \Phi(-b_1),}}
\end{array}
\end{equation}
where the $S_0 \rightarrow \tilde{S}_0$ adjustment should be done in $b_1$ and $b_2$ coefficients as well.

In addition, \cite{Frishling2002} reports another, also quite conventional, approximation to account for discrete dividends - instead of changing $S_0$ in Eq. (\ref{BS_conven}) one can adjust the strike $K$ instead:
\begin{equation} \label{BS_strike}
\begin{array}{l}
{\displaystyle
{C=S_0 \Phi(b_1)- \tilde{K} \exp{(-r T)} \Phi(b_2),}}
\\
\\
{\displaystyle
{P = \tilde{K} \exp{(-rT)} \Phi(b_2) - S_0 \Phi(-b_1),}}
\end{array}
\end{equation}
where $\displaystyle {\tilde{K} = K + \sum_{0<t_i\le T}{d_i \exp{(r (T - t_i))}}}$ and, again, the same adjustment is done in $b_1$ and $b_2$ coefficients. Below we will refer to this approximation as the vanilla strike adjustment approximation or simply \emph{strike approximation}.

After providing Eqs. (\ref{BS_spot}) and (\ref{BS_strike}) approximations \cite{Frishling2002} stops outlining analytical results and points out that both of these results may significantly differ from numerical CN modelling by either underpricing [Eq. (\ref{BS_spot})] or overpricing them [Eq. (\ref{BS_strike})] if the same volatility is used and dividends are paid as at ex-dividend days. This is because for vanilla spot and vanilla strike approximations the stock process (\ref{stockprocess}) has been strictly increased or decreased relative to the strike, but the volatility $\sigma$ has not been altered to reflect that.

\cite{BosV2002} agrees with these conclusions of \cite{Frishling2002}, but in addition suggest another BS approximation (which it also supports with reasonably involved theoretical backing):

\begin{equation} \label{BS_hybrid}
\begin{array}{l}
{\displaystyle
{C=\bar{S}_0 \Phi(b_1)- \bar{K} \exp{(-r T)} \Phi(b_2),}}
\\
\\
{\displaystyle
{P = \bar{K} \exp{(-rT)} \Phi(b_2) - \bar{S}_0 \Phi(-b_1),}}
\end{array}
\end{equation}
where $\bar{S}_0 = S_0 - D_S$ and $\bar{K} = K + D_K \exp{(r T)}$ where, in turn, $\displaystyle {D_K \equiv \sum_{0<t_i\le T}{\frac{t_i}{T} d_i \exp{(-r t_i)}}}$ and $\displaystyle {D_S \equiv \sum_{0<t_i\le T}{\frac{T - t_i}{T} d_i \exp{(-r t_i)}}} = D - D_K$. The approximation (\ref{BS_hybrid}) was claimed to be in a good agreement with CN-based numerical results, but no details for numerical modelling were provided. Below we will refer to this approximation as \emph{hybrid approximation}.

 \cite{BenederW2001} uses a different approach for their approximation. They start by noting that if the local volatility of a stock process with discrete dividends is constant, then the corresponding process without dividend-induced jumps [where $S_0$ is adjusted as $\displaystyle {\tilde{S}_0 = S_0 - \sum_{0<t_i\le T}{d_i \exp{(-r t_i)}}}$] should have non-constant local volatilities:

\begin{equation} \label{local_adjustment}
\tilde{\sigma}_S(S,D,t) = \sigma(T) \frac{S}{S-D_j^{(S)}},
\end{equation}
where $\displaystyle {D_j^{(S)} =  \sum_{i = j(t)}^{N}{d_i \exp{(-r t_i)}}}$ and where, in turn, $N$ is the number of dividend payouts within the $(0,T)$ interval and the sum only includes those dividend payments that occur after time $t$ [with $j(t)$ being the index number of the first dividend payment occurring at or after $t$ point of time: $t_{j-1} < t \le t_j$]. For example: $D_j^{(S)}(0) = D_1^{(S)} \equiv D$.

Furthermore, the corresponding variance $\tilde{\sigma}_S^2(S,D,t)$ may be averaged on the $(0,T)$ interval to obtain:
\begin{equation} \label{VA_spot}
\bar{\sigma}_S = \sigma \sqrt{ \left(\frac{S}{S-D_1^{(S)}}\right)^2 \frac{t_1}{T} +
\sum_{1<j<N}\left(\frac{S}{S-D_j^{(S)}}\right)^2 \frac{t_j - t_{j-1}}{T} + \frac{T - t_N}{T}} \equiv
\sigma (1 + \varepsilon_S),
\end{equation}
where $t_N$ is the timing of the last dividend payment within the $(0,T)$ interval.

To use Eq. (\ref{VA_spot}) one should simply replace $\sigma$  with $\bar{\sigma}_S$  in the spot approximation system given by Eqs. (\ref{BS_spot}). Below we will refer to this approximation as spot volatility adjusted approximation (or spot VA approximation). Note that if dividend payouts only occur in close vicinity of $t = 0$, then $\bar{\sigma}_S \approx \sigma$  with a high degree of accuracy.

It is interesting to note that \cite{BosGS2003}, independently of \cite{BenederW2001}, has suggested a more rigorously backed version of $\bar{\sigma}_S$ volatility adjustment, which, according to the analysis of \cite{HaugHL2003,VellekoopN2006}, provides slightly better agreements with numerical results. Here we will limit ourselves to analysing the spot VA approximation version of \cite{BenederW2001} only (in order not to be overwhelmed by the classification of minimally different methods). However, it is important to note that \cite{BosGS2003}, in contrast to \cite{BenederW2001}, has also outlined the possibility of yet another version of VA approximation: strike VA approximation. Note, that the authors of \cite{BosGS2003} have not provided a final explicit expression for this new approach, only mentioning that this can be ``easily done''. Thus we have chosen to present strike VA approximation by a straightforward generalisation of Eqs. (\ref{local_adjustment}) and (\ref{VA_spot}) which has a much simpler form in comparison to the corresponding expressions of \cite{BosGS2003}. We note, however, that one can obtain \cite{BosGS2003}-style strike VA volatility adjustment expression by modifying our Appendix A formulas (setting $\alpha_i = 0$ in the last formula of Appendix A).

Similar to the vanilla spot approximation generalisation of \cite{BenederW2001} described above, a vanilla strike approximation [given by Eqs. (\ref{BS_strike}) with a strike adjustment of
$\displaystyle {\tilde{K} = K + \sum_{0<t_i\le T}{d_i \exp{(-r (t_i - T))}}}$] can be chosen as a basis for a further volatility adjustment. Then, using arguments similar to spot VA case, we can rewrite Eq. (\ref{local_adjustment}) as:
\begin{equation} \label{local_adjustment2}
\tilde{\sigma}_K(S,D,t) = \sigma(T) \frac{S}{S+D_j^{(K)}},
\end{equation}
where $\displaystyle { D_j^{(K)} =  \sum_{i = 1}^{j(t)}{d_i \exp{(-r t_i)}}}$ and where, in turn, the sum only includes those dividend payments that occur before time  $t$ [with $j(t)$  being the index number of the last dividend payment occurring at or before $t$ point of time: $t_j \le t < t_{j+1}$. For example:
$D_j^{(K)}(T) = D_N^{(K)} \equiv D$.

Furthermore, the corresponding variance $\tilde{\sigma}_K^2(S,D,t)$  may be averaged on the $(0,T)$ interval to obtain:
\begin{equation} \label{VA_strike}
\bar{\sigma}_K = \sigma \sqrt{ \frac{t_1}{T}  +
\sum_{1<j<N}\left(\frac{S}{S+D_j^{(K)}}\right)^2 \frac{t_j - t_{j-1}}{T} + \left(\frac{S}{S+D_N^{(K)}}\right)^2 \frac{T-t_N}{T}} \equiv
\sigma (1 - \varepsilon_K),
\end{equation}
where $t_N$ is the timing of the last dividend payment within the $(0,T)$ interval.

To use Eq. (\ref{VA_strike}) one should simply replace $\sigma$  with $\bar{\sigma}_K$  in the strike approximation system given by Eqs. (\ref{BS_strike}) Below we will refer to this new approximation as strike VA approximation. Note that if dividend payouts only occur in close vicinity of $t=T$, then $\bar{\sigma}_K \approx \sigma$  with a high degree of accuracy.

It is important that both spot VA and strike VA approximations essentially represent slightly different versions of the same general perturbation approach with different zeroth order approximations (vanilla spot and vanilla strike approximations respectively), and with different corresponding small parameters $\displaystyle {\varepsilon_S = \frac{D/S}{1-D/S}}$ (for spot VA) and $\displaystyle {\varepsilon_K = \frac{D/S}{1+D/S}}$ (for strike VA). In the case of $D \rightarrow S$,  one can expect strike VA approximation to outperform spot VA approximation because the former tends to have a much smaller $\varepsilon$  ($\varepsilon_K \rightarrow 0.5$  versus $\varepsilon_S \rightarrow \infty$). However even $\varepsilon_K \approx 0.5$  is likely to be too large of a perturbation parameter value to provide a high quality approximation.

Before ending this subsection and moving to the description of our numerical benchmark technique, we would like to mention another interesting recent approach based on asymptotically exact, Taylor-like series expansion (TE approach of \cite{VeigaW2009}). We are grateful to the authors of this work for sharing their codes with us, allowing a quick reproduction of their results which we will compare with other methods in Section \ref{results}.

It is also important to point out that Put and Call values given by any of the approaches presented above satisfy the famous Put-Call Parity relation:
\begin{equation} \label{parity_BS}
K \exp(-r T) - P = S_0  \exp(-q T) - C,
\end{equation}
(see e.g. \cite{Hull2006} for derivation details).

\subsection{Review of Crank-Nicolson schemes}\label{CN}

Our aim is to justify our choice of a benchmark direct numerical modelling, which is based on a rather conventional version of the Crank-Nicolson (CN) scheme. However, even here we face some uncertainties about the details of its numerical implementation, which need to be resolved.
The conventional Black-Scholes equation governing the option price dynamics is given by
\begin{equation} \label{BS}
\frac{\partial V}{\partial t} - r V + r S \frac{\partial V}{\partial S} +
\frac{1}{2} \sigma^2 S^2 \frac{\partial^2 V}{\partial S^2} = 0,
\end{equation}
where the asset price (spot) is $S$, the (flat) volatility is $\sigma$, the time is $t$, and the term structure of forward interest rates is $r(t)$ [for simplicity we shall only consider flat interest rates $r(t) = r$].  This equation is integrated backward from the expiry date $T$ on a finite difference grid of asset values $S_i$ (where $ \displaystyle{i = 1,  2,  3, \ldots ,  N}$) using a CN finite difference scheme.

The CN method is well known in financial applications and beyond. Good reviews of the methods and its variations can be found e.g. in \cite{WilmottHD1995,TavellaR2000}. Essentially CN finite-difference systems are solved, using either a direct LU method (typically used to European options pricing), or a so called projected successive over-relaxation (PSOR) method (typically used for American option pricing). Most of the existing literature on CN is devoted to American pricing option discussions (see, e.g. \cite{ZhaoDC2007,EhrhardtM2008})  while substantially fewer works focus on European options (see, e.g. \cite{SivenSP2009} for a relatively rare example).

In general, CN algorithms for American options are significantly more complex, than their European counterparts and involve a so-called free-boundary problem formulation (see e.g. \cite{WilmottHD1995,TavellaR2000} for details). However, in this work we concentrate on European options and will only need American CN solver for the calibration/clarification of our choice of boundary conditions of its European analog. Thus, we choose not to implement a free-boundary solver, but instead, for Puts only, use a fixed boundary American CN PSOR scheme with:
\begin{equation} \label{Boundary_AM}
\begin{array}{l}
{\displaystyle
{V_N(0 \le t \le T) = 0,}}
\\
\\
{\displaystyle
{V_1(0 \le t \le T) = K - S_1}}
\end{array}
\end{equation}
boundary conditions, where  index $N$ corresponds to the upper and index $1$ to the lower boundary of the discretisation grid. It is relatively easy to justify this choice, because it is well known that an early exercise is always the preferred option for American Puts if spot falls to close to zero levels (see e.g. \cite{Hull2006} for a related discussion).

For the completeness of our presentation we also state the (standard) initial conditions (which are the same both for American and European CN solvers):
\begin{equation} \label{put_IC}
V(T) = \max(K-S_i, 0),
\end{equation}
for Puts [and  $V(T) = \max(S_i-K, 0)$ for Calls], where $1 \le i \le N$.

Although the choice of boundary conditions for European CN schemes is far from being straightforward, typically it is not discussed in the literature. We start our analysis of boundary conditions by reiterating some core results of \cite{HaugHL2003}, where discrete dividend option pricing is described in a methodologically consistent fashion. Among other important things, \cite{HaugHL2003} reports an observation that put option pricing is dependent on the dividend policy, i.e. an assumption of how much the company will pay if its spot price falls very low and formally $S(t_i^-) - d_i <0$, where $d_i$ is the forecasted dividend payout at   $t_i$ and the minus superscript refers to the time instantaneously before $t_i$. Two policies are suggested: a Survivor policy [$d_i \rightarrow \tilde{d}_i = 0$] and a Liquidator policy [$d_i \rightarrow \tilde{d}_i = S(t_i^-)$]. Although in practice, the Survivor policy is more prudent during troubled times, for modelling purposes the Liquidator policy is often adopted explicitly or implicitly, because it provides the dividend $\tilde{d}_i(S)$ as a continuous function in $S$:
\begin{equation} \label{liquidator}
\tilde{d}_i^{(l)}(S(t_i^-))= \min(S(t_i^-),d_i),
\end{equation}
whereas the survivor policy has a discontinuity:
\begin{equation} \label{survivor}
\tilde{d}_i^{(s)}(S(t_i^-))= d_i H(d_i -S(t_i^-)),
\end{equation}
where $H(a)$ is the Heaviside step function [$H(a < 0) = 0$ , $H(a \ge 0) = 1$].

It is quite obvious now that a special care should be taken in both adjustments of the grid at discrete dividend payout days (and, for European options, boundary conditions as well).

As we work backwards on the grid through a policy-dependant dividend payment $\tilde{d}_i$ at time $t_i$, the asset price rises by the amount of the dividend.  Thus, we have to introduce the shift $S(t_i^-) = S(t_i^+) + \tilde{d}_i$, which, due to the continuity of derivative price requirement, in turn,  leads either to the equation
\begin{equation} \label{VshiftEURO}
V^-[S, t_i^-] = V^+[S - \tilde{d}_i, t_i^+],
\end{equation}
(for European CN scheme) or to its American counterpart version
\begin{equation} \label{VshiftAM}
V^-[S, t_i^-] = \max (V^+[S - \tilde{d}_i, t_i^+], K-S(t_i^+)),
\end{equation}
for which we only present the Put version.

European CN boundary condition formulation for stocks with discrete dividends is far less obvious and differs significantly from its continuous dividend analogue (given e.g. by \cite{SivenSP2009}).  We will illustrate the problem by looking at the example of the liquidator policy (\ref{liquidator}).
It is clear that we have to adjust the standard continuous dividend version of boundary conditions [which have the form of $V_N(0 \le t \le T) = 0$ and $V_1(0 \le t \le T) = K \exp(-r (T-t)) - S_1 \exp (-q (T-t))$, where $q$ is a continuous dividend yield] into an expression which includes some discrete dividend-related negative asset price exclusion constraints. However, it may seem that there are a few possible ways to do this. Indeed, after the construction of a time dependent discounted dividend sum in a $(t,T)$ interval $\displaystyle {\tilde{D}(t) = \sum_{t \le t_i \le T} d_i \exp{(-r (t_i-t))}}$  and recalling the different approaches from the previous subsection, one may suggest at least 3 different versions of a lower boundary condition:
\begin{equation} \label{SpotBC}
V_1(0 \le t \le T) = K \exp (-r (T-t)) - \max(S_1 - \tilde{D}(t),0),
\end{equation}
\begin{equation} \label{StrikeBC}
V_1(0 \le t \le T) = [K \exp (-r (T-t)) + \tilde{D}(t)] - S_1,
\end{equation}
\begin{equation} \label{HybridBC}
V_1(0 \le t \le T) = [K \exp (-r (T-t)) + \tilde{D}_K(t)] - \max(S_1 - \tilde{D}_S(t),0),
\end{equation}
where $\displaystyle {\tilde{D}_K(t) \equiv \sum_{t \le t_i \le T} \frac{t_i d_i \exp{(-r (t_i-t))}}{T}}$ and $\displaystyle {\tilde{D}_S(t) \equiv \sum_{t \le t_i \le T} \frac{(T-t_i) d_i \exp{(-r (t_i-t))}}{T}}$.

These versions may be interpreted as the corresponding spot, strike and hybrid BS approximations from the previous subsection, respectively. By choosing $S_1$ close to zero, we can reduce Eqs. (\ref{SpotBC}-\ref{HybridBC}) to a simpler representation $V_1(0 \le t \le T) = K \exp (-r (T-t)) + \bar{D}$, where $\bar{D}$  is $0$ or $\tilde{D}$ or $\tilde{D}_K$, respectively. Without further analysis it is not obvious which of Eqs. (\ref{SpotBC}-\ref{HybridBC}) is preferable with Eqs. (\ref{SpotBC}) and (\ref{StrikeBC}) being particular strong candidates. [Note that, in contrast to Puts, Calls are indifferent to Eqs. (\ref{SpotBC}-\ref{HybridBC})-like issue, because the function $\max(S_N - \tilde{D}(t),0)$ never returns zero if $S_N$ is chosen at an appropriately high level and the upper boundary condition can be straightforwardly chosen in the form of $V_N(0 \le t \le T) = S_N - \tilde{D} - K \exp (-r (T-t)) $]. Below we adopt a simple approach of running a modelling test to comparing our 3 versions of European CN schemes for Puts [with boundary conditions (\ref{SpotBC}-\ref{HybridBC}) respectively] with the corresponding American scheme.

\begin{figure}[h]
\begin{center}
\resizebox*{11cm}{!}{\includegraphics{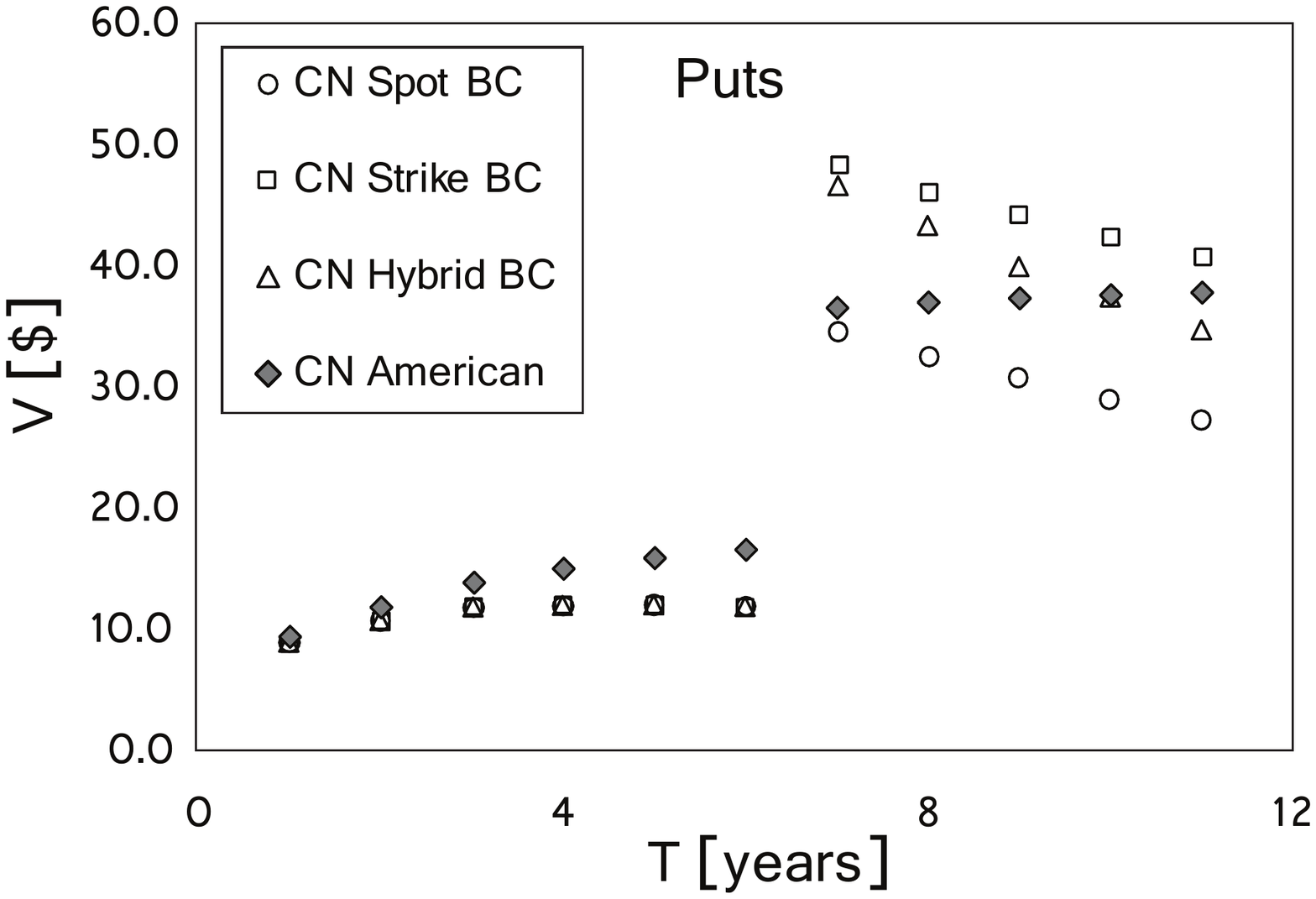}}%
\caption{Comparison of Put results given by different versions of European CN scheme (with different choice of boundary conditions) and the corresponding American CN scheme (see text for details)}%
\label{Fig1}
\end{center}
\end{figure}r

For demonstration purposes we will present our findings using a sample family of Put options: we apply flat discounting rate $r=6\%$ and flat volatility $\sigma = 30 \%$, $S_0 = \$100.0$, $K = \$100.0$ and maturity terms $T$ ranging from 1.0 to 11.00 years (with a one year increment). In addition we choose a single forecasted dividend payout of $d_1 = \$70.00$  at the $t_1 = 6.5$ year mark. The results are presented in Fig. \ref{Fig1}. One can easily see that until the first (and only) dividend payout day, the boundary conditions given by Eqs. (\ref{SpotBC}-\ref{HybridBC}) coincide and lead to identical CN results. After the $t_1 = 6.5$ year mark the situation changes and substantial differences appear. Most importantly we note that only the boundary conditions given by Eq. (\ref{SpotBC}) lead to the result which is consistent with the corresponding American Put result. Indeed, European option prices should always stay smaller than their American counterparts - a behaviour displayed by CN scheme with boundary condition (\ref{SpotBC}), but not with (\ref{StrikeBC}) or (\ref{HybridBC}).

To remove any remaining doubts we also compared our CN results [with boundary condition (\ref{SpotBC})] against a standard Monte Carlo (MC) scheme (see e.g. \cite{Wilmott2006} for a description) with Liquidator-style path adjustment at ex-dividend days. Our antithetic MC code with $10^6$ paths closely matched CN results (less than $0.05\%$ discrepancies for each of 11 Puts).

Again it is important to reiterate that the boundary condition uncertainty for European CN schemes is only present for Puts. For Calls, any choice of boundary conditions (with or without $\max$  function and with or without dividend split) leads to identical results, which is in line with an observation of \cite{HaugHL2003}, that a company dividend policy may only affect Puts, but not Calls.

\subsection{Qualitative comparison of CN numerics with analytic results}

\begin{figure}[h]
\begin{center}
\resizebox*{11cm}{!}{\includegraphics{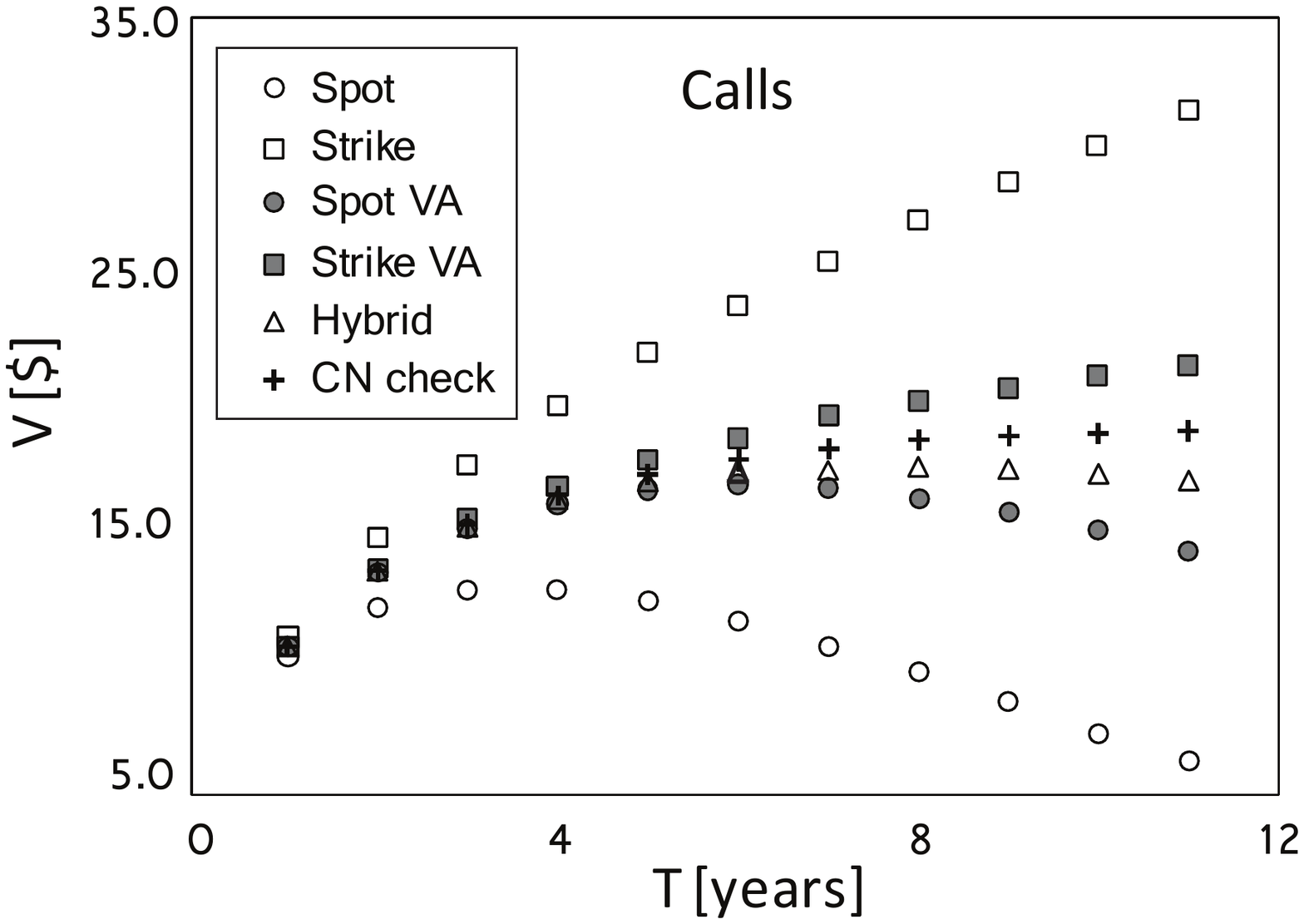}}%
\caption{Comparison of results given by different analytical approaches for a multiple dividend family of Calls (see text of the paper for details; corresponding Put results are presented in Fig. \ref{Fig3}). Spot approximation is given by Eqs. (\ref{BS_spot}); Strike approximation - by Eqs. (\ref{BS_strike}); Hybrid approximation - by Eqs. (\ref{BS_hybrid}); VA Spot approximation - by Eqs. (\ref{BS_spot}) with adjusted volatility given by Eq. (\ref{VA_spot}); VA Strike approximation - by Eqs. (\ref{BS_strike}) with adjusted volatility given by Eq. (\ref{VA_strike}); and finally CN numerical results - by modelling Eq. (\ref{BS}) with initial conditions (\ref{put_IC}) and boundary conditions (\ref{SpotBC}). }%
\label{Fig2}
\end{center}
\end{figure}

\begin{figure}[h]
\begin{center}
\resizebox*{11cm}{!}{\includegraphics{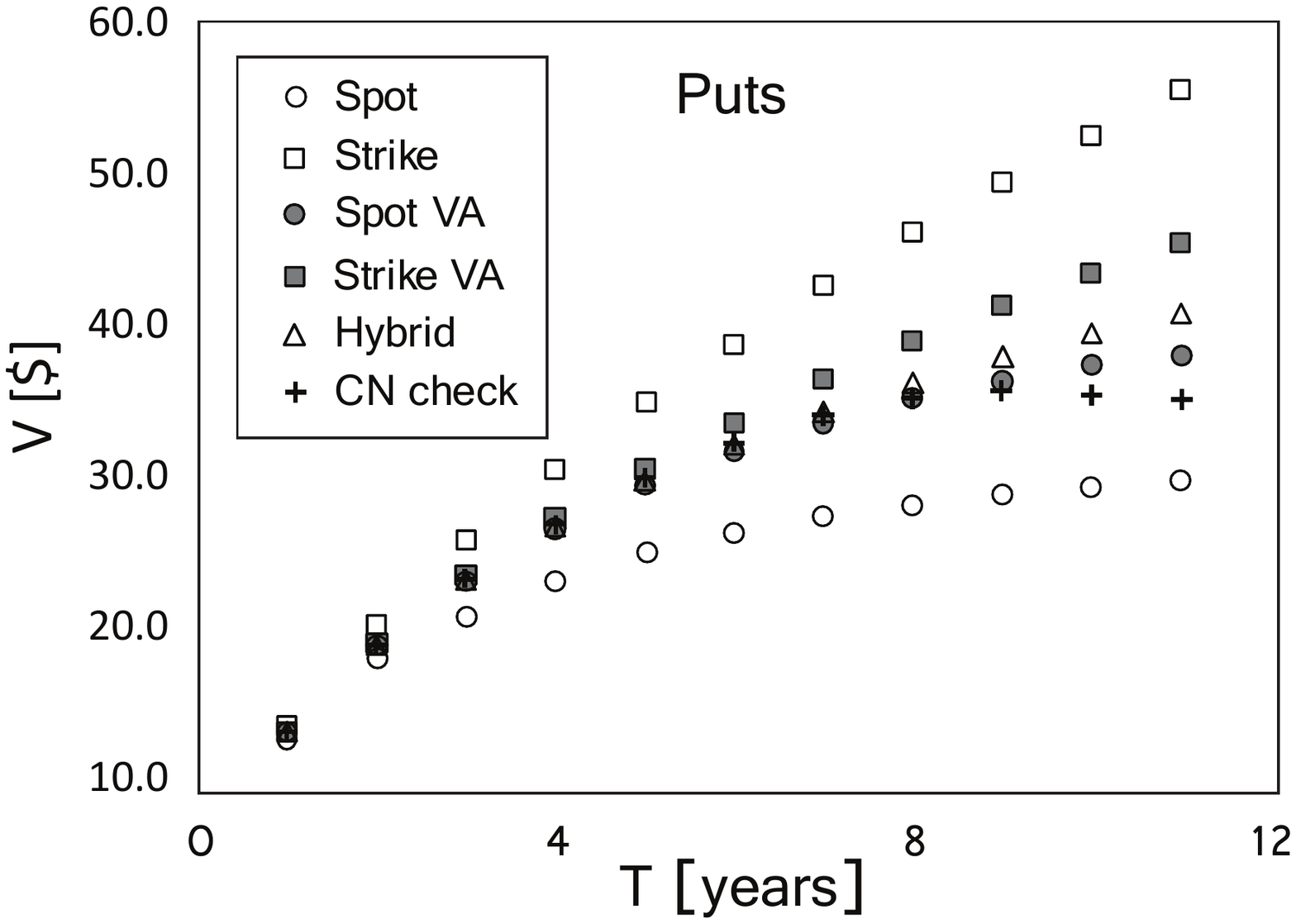}}%
\caption{Comparison of results given by different analytical approaches for a multiple dividend family of Puts (see text of the paper for details; corresponding Call results are presented in Fig. \ref{Fig2}). }%
\label{Fig3}
\end{center}
\end{figure}

Now we are in the position to do an initial (qualitative) comparison of different analytic methods with a benchmark numerical method [CN scheme with spot-type boundary conditions (\ref{SpotBC})]. We will present this using a sample family of European Calls and Puts. We apply flat discounting rate $r = 6 \%$ and flat implied volatility $\sigma = 30\%$ (we will also use as flat volatility for CN modelling), $S_0 = \$100.0$, $K = \$100.0$ and maturity terms $T$ ranging from 1.0 to 11.00 years (with a one year increment). The family also has regular annual dividend payouts of $d_i = \$9$  at the $t_i = 0.5, 1.5, 2.5, \ldots, 10.5$  year marks, arguably  representing a real-life situation of e.g options on a profitable,  but limited life-span (e.g. due to commodity reserves constraints) mining stock. We also note that our test cases have substantially higher dividends, than those considered by e.g. \cite{HaugHL2003,VellekoopN2006}. We intentionally aimed to push all approximations to their applicability limits to provide an easily observable evidence of their agreement (or disagreement) with numerics. Note that the corresponding numerical result of best performing approximations of Figs. \ref{Fig2} and \ref{Fig3} are also presented in Section \ref{results} in a tabular format. Here and below we used $\Delta t = 0.05$, $S_{\max} = 500$, $S_{\min} = 0$ and $\Delta S = 1.25$ for our CN modelling.

Several conclusions can be drawn from analysis  Figs. \ref{Fig2} and \ref{Fig3}:
\begin{itemize}
{\item None of presented analytical approximations is good for longer dated options (both Calls and Puts) with simpler Spot and Strike approaches being particulary different from the benchmark numerical results.
}
{\item Interestingly, the numerical results for Calls and Puts demonstrate qualitatively different trends in their disagreements with the best of analytic approximations (Hybrid and Strike VA): Calls in general follow Strike VA and Hybrid averaged trend, whereas longer-dated Puts deviate significantly below both Hybrid and Strike VA curves.
}
{\item
Qualitatively different deviations of numerical curves for Puts and Calls from their analytic counterparts implies a violation of the Put-Call parity relation (\ref{parity_BS}): if parity was preserved, then the deviation of numerics from any one of the analytic approximations presented above would be the same for Puts as it is for the corresponding Calls.
}
\end{itemize}

Thus we can formulate the two major issues to be addressed by this paper:
\begin{enumerate}
{\item Is the observed parity violation phenomenon real and if ``yes'' then how it can be explained?}
{\item Can a simple BS-type approximation better matching the benchmark numerical results be developed for pricing European Calls and Puts for equities with large discrete dividends?}
\end{enumerate}

The following two subsections will consequently address both of these questions.

\subsection{Parity violation phenomenon}

 According to our knowledge the parity violation phenomenon was first described by \cite{HaugHL2003}, where, however, the term ``violation'' was not used per se - the authors have derived and presented a modified form of the parity relation (see Eq. (8) of \cite{HaugHL2003}). Due to the lack of stress on the changed form of the parity relation, the important difference between the modified relation and the conventional one may have been missed by a few of readers of \cite{HaugHL2003}, including even those referring to this paper in their own subsequent publications. Thus, in our opinion, it is important to revisit the issue and clearly reveal reasons of the parity violation phenomenon.

Let us consider a stock process $S(t)$ with a single deterministic dividend payout $D$ at time $t = t_D$ and suppose for simplicity that $r=0$. If the dividend payout is guaranteed in timing and value (e.g. by holding the corresponding amount in an escrow account), then by construction for every observable stock price $S(T)$ $(S(T) > 0)$ we have the standard parity relation at payout:
\begin{equation} \label{parity_at_payout}
C(T) - P(T) = S(T) - K.
\end{equation}
If $D(t_D) = {\rm const}$ as discussed, then it is easy to construct a portfolio at $t=0$ that has no intermediate net cashflows, and matches the right-hand side of the equation (\ref{parity_at_payout}) at time $T$. One needs to buy the stock $S$, short $K$ (to be paid back at $t=T$) and short $D$ (to be paid back at $t=T_D$). The net initial cost of such a portfolio is $S(0) - K - D$.  Importantly the time $t_D$ the payments net out exactly ($D - D = 0$), and the cash flow at $t = T$ is $S(T)-K$, as we intended. Thus, we have satisfied both conditions used to derive the standard parity relation (\ref{parity_BS}), i. e. we constructed the portfolio with payout (\ref{parity_at_payout}) at $t=T$ and demonstrated the absence of non-zero cashflows in the interval $0<t<T$. Accordingly the modified parity relation stays:
\begin{equation} \label{parity_at_origin}
C(0) - P(0) = S(0) - K - D,
\end{equation}
which (after an extension of our analysis to $r \neq 0$ case) is equivalent to Eq. (\ref{parity_BS}).

However, with the Liquidator/Survivor dividend policies (described in Subsection \ref{CN}) in place the argument above does not work because there is a non-zero probability that the payments at $t=t_D$ will not cancel. Indeed, if $S(t_D) < D$, then we will not receive $D$ amount needed to cover the corresponding short position, but only either $S(t_D)$ (if Liquidator policy is in place) or $0$ (if Survivor policy is chosen). Thus, the standard parity relation (\ref{parity_at_origin}) fails. However, the true put-call parity can be found by calculating the expected
missing cashflows (e.g. by the risk-neutral valuation technique; see \cite{Hull2006} for details). But up to the dividend date, the stock price follows a standard Geometric Brownian motion, so that the original Black-Scholes formulas may be applied. For a single dividend case with Liquidator policy, such calculations result in parity violation value given by the standard BS Put pricing formula
\begin{equation} \label{singleL}
\Delta P = D \exp{(-rT_D)} \Phi(-b_2) -S_0 \Phi(-b_1),
\end{equation}
where $K$ is replaced by a non-discounted single dividend value $D$, $T_D$ is the single dividend payout timing, and $D$ and $T_D$ also replacing $K$ and $T$ in the expressions for $b_1$ and $b_2$. Note, that for a single dividend case $\Delta P$ formula does not require any volatility adjustments.

Similar calculations show that for a single dividend case with Survivor policy, the corresponding result is given by
\begin{equation} \label{singleS}
\Delta P = D \exp{(-rT_D)} \Phi(-b_2),
\end{equation}
where again no volatility adjustments are needed.

For multi-dividend cases, no exact parity violation expressions are available and we are not aware of any previously reported analytic approximations.

\subsection{New analytic Black-Scholes-type approximations for calls and puts}

All analytical approaches described in this work so far can be divided into three groups:
\begin{enumerate}
\item{Simple heuristic approaches attempting to adjust some of BS formula parameters, but not the volatility to take the effect of discrete dividends into account (Spot, Strike and Hybrid approaches)}
\item{Approaches attempting to adjust some of BS formula parameters, and \emph{then} adjust the volatility to fine-tune the correction (Spot VA and Strike VA)}
\item{Asymptotic expansion approach (TE approach)}
\end{enumerate}
It is quite easy to notice that, although there are \emph{three} simple methods of type 1 (Spot, Strike and Hybrid approaches), only \emph{two} corresponding volatility-adjusted approaches have been reported (Spot VA and Strike VA).
Moreover, these volatility adjusted approaches used Spot and Strike approximations as their starting points, which represent a rather poor choice. It is more natural to take Hybrid approximation as the starting point. Indeed, our Figs. \ref{Fig2} and \ref{Fig3} clear show that the hybrid approximation [Eqs. (\ref{BS_hybrid})] agrees reasonably well with numerics even without any further volatility adjustment. Now we are in a strong position to take one step further and construct the hybrid VA approximation in a most straightforward fashion: (i) we start with Eqs. (\ref{BS_hybrid}) where the dividends are split into $D_S$ and $D_K$ parts ($D = D_S + D_K$); (ii) then we calculate volatility adjustments as defined by Eqs. (\ref{VA_spot}) and (\ref{VA_strike}), but with $\displaystyle {D_S \equiv \sum_{0<t_i\le T}{\frac{T - t_i}{T} d_i \exp{(-r t_i)}}}$ and $\displaystyle {D_K \equiv \sum_{0<t_i\le T}{\frac{t_i}{T} d_i \exp{(-r t_i)}}}$, i.e. we calculate volatility adjustments \emph{independently} for $D_S$ and $D_K$ parts of $D$ [to distinguish newly obtained $\varepsilon_S$ and $\varepsilon_K$ from their counterparts for spot and strike VA approximations we will denote them as $\varepsilon_S^{(h)}$ and $\varepsilon_K^{(h)}$ below]; (iii) finally we calculate a new volatility adjustment coefficient as a direct product of $(1+\varepsilon_S^{(h)})$ and $(1-\varepsilon_K^{(h)})$:
\begin{equation} \label{VA_hybrid}
\bar{\sigma}_H = \sigma (1+\varepsilon_S^{(h)}) (1-\varepsilon_K^{(h)}) \equiv \sigma (1-\varepsilon_H).
\end{equation}
Note that the new perturbation parameter $\varepsilon_H$ is typically significantly smaller than $\varepsilon_S$ or   $\varepsilon_K$ because of the dividend split (typically for real-life situations $D_S \approx D_K \approx D/2$) and also because the oppositely signed contributions of $\varepsilon_S^{(h)}$ and $\varepsilon_K^{(h)}$ in $(1+\varepsilon_S^{(h)}) (1-\varepsilon_K^{(h)})$ product.

To summarise: Hybrid approximation [given by Eqs. (\ref{BS_hybrid})] extended by a volatility adjustment of $\sigma \rightarrow \bar{\sigma}_H$  [given by Eq. (\ref{VA_hybrid})] explicitly describes our {\em new approximation}, which we refer to as Hybrid VA approximation. We further note, that Hybrid VA approximation for Calls is final, whereas Hybrid VA approximation for Puts needs further adjustment to account for the parity violation phenomenon.

For a single dividend case parity violation adjustment (PA) is given by Eq. (\ref{singleL}) [Liquidator dividend policy - effective Put] and Eq. (\ref{singleS}) [Survivor dividend policy - effective digital Put]. Thus we can simply calculate parity violation adjusted Puts as given by the second of  Eqs. (\ref{BS_hybrid}) with an extra volatility adjustment according to Eq. (\ref{VA_hybrid}) and then subtract parity violation value given by Eq. (\ref{singleL}) or Eq. (\ref{singleS}). We refer to this approach as Hybrid VAPA (Hybrid volatility adjusted parity violation adjusted) approach. We remind that parity violation adjustment is only relevant for Puts, but not for Calls.

For multi-dividend case no explicit formula for PA is available. However, in the spirit of this paper we may suggest calculating multi-dividend PA as an effective Puts (similarly to a single dividend case). Then we can use e.g. already developed Hybrid VA approximation for the valuation of these effective Puts. Indeed, one can almost effortlessly generalise Eq. (\ref{singleL}) into:
\begin{equation} \label{multiL}
\Delta P^{(L)} = P_{{\rm eff}}^{(L)} = \bar{D}_L \exp{(-rT_D)}  \Phi(-b_2) -\bar{S}_0 \Phi(-b_1),
\end{equation}
where $D_L$ is the last dividend, $T_D$  is time of the last dividend payout, $\bar{S}_0 = S_0 - \tilde{D}_S$ and $\bar{D}_L = D_L + \tilde{D}_K \exp(r T_D)$  where, in turn, $\displaystyle{\tilde{D}_S \equiv \sum_{0<t_i< T_D}{\frac{T_D - t_i}{T_D} d_i \exp{(-r t_i)}}}$ and $\displaystyle {\tilde{D}_K \equiv \sum_{0<t_i< T_D}{\frac{t_i}{T_D} d_i \exp{(-r t_i)}}}$. Note that the sums $\tilde{D}_S$ and $\tilde{D}_K$ both exclude the last dividend payout because it is already taken into account by being an effective unadjusted strike [$K \rightarrow D_L$]. In addition, volatility in Eq. (\ref{multiL}) is also adjusted according to the standard Hybrid VA approximation recipe [i.e. the correction is given by Eq. (\ref{VA_hybrid}), where now $T \rightarrow T_D$ and all summations in Eqs. (\ref{VA_spot}) and (\ref{VA_strike}) are on the $(0, T_D)$ interval and do not include the final dividend].

Multi-dividend generalisation of Eq. (\ref{singleS}) may be done in a similar fashion, but with one important difference -- the effective Put represented by Eq. (\ref{multiL}) does not include further recursive parity violation adjustments, whereas the generalisation of Eq. (\ref{singleS}) should include these recursive terms: $\displaystyle{\Delta P^{(S)}(t_N) =\sum_{i= 1}^N (-1)^{N-i} P_{{\rm eff}}^{(S)}(t_i)}$, where $t_i$ are ex-dividend days and $t_N$ is the timing of the last dividend [within the $(0, T)$ interval]. Indeed, for the Liquidator case all paths of process (\ref{stockprocess}) which collapse to zero at ex-dividend day $t_i$ will continue staying at zero and do not have any chance of path recovery above $d_{i+1}$ for the next ex-dividend date $t_{i+1}$, whereas for the Survivor case such a recovery is always possible.

We stress that the main idea of our PA approach here is to view it as effective multi-dividend Put (vanilla for Liquidator policy or digital for Survivor policy). The application of Hybrid VA algorithm for calculation of such an effective Put value is an extra step. In fact, we can utilise any other approximation for this purpose (with a preference obviously given to higher accuracy algorithms).

In the next Section we comprehensively compare the best of previously reported analytic approximation for Calls and Puts (Spot VA, Strike VA, Hybrid and TE approaches) with the corresponding CN numerics and the newly developed Hybrid VA and Hybrid VAPA analytic approximations.

We also note that the newly developed Hybrid VA approach (and its parity violation adjusted extension for Puts: Hybrid VAPA approach) are essentially based on rather heuristic ideas of \cite{BosV2002} and \cite{BenederW2001}. The later approach does not allow calculation of higher order corrections of volatility adjustment. Such corrections can be calculated by extending the result of \cite{BosGS2003}, who suggested a much more complex, but also more rigorous volatility adjustment approach. For the completeness of our analysis we extended the analysis of  \cite{BosGS2003} calculating the corresponding Hybrid VA-2 approximation formulas (second [more rigorous and more complex] version of Hybrid VA method). All relevant formulas and derivation details can be found in Appendix A. Hybrid VA-2 results are also compared with numerics and other analytical approximations in the next Section.

\section{Results: final comparison of different analytic approaches with CN numerics} \label{results}

We present our findings using two sample families of European options with large discrete dividends. For both families we apply flat discounting rate $r = 6 \%$ and flat volatility $\sigma = 30\%$ (which we will also use as flat rate curve and volatility for CN modelling), $S_0 = \$100.0$, $K = \$100.0$ and maturity terms $T$ ranging from 1.0 to 11.00 years (with a one year increment). First family has only one forecasted dividend payout of $d_1 = 50.0$ at $t_1 = 364/365$ mark (i.e. one day before expiry of shortest term option (with $T=1$); thus we can observe a clear transition between the limit where the only dividend is close to expiry [for $T=1$ option] and the opposite limit where the only dividend is relatively close to the origin [e.g. for $T = 11$ option]). The second family is the same as the presented in Figs. \ref{Fig2} and \ref{Fig3}: it has regular annual dividend payouts of $d_i = 9$ at $t_i = 0.5, 1.5, 2.5, \ldots, 10.5$  year marks. First family is used essentially for "stress-testing" of our analysed analytical models, whereas the second one, as we have already mentioned, may represent a real-life situation.

\begin{table}[h]
  \tbl{Comparison of different analytic methods with CN numerical results for a single-dividend family of Calls. First column shows CN numerical results. In consequent columns different analytic approximations and their corresponding relative differences with respect to CN numerics are presented.}
{\begin{tabular}{|c|c|c|c|c|c|c|c|c|c|c|c|c|c|} \hline
& & & & & & & & & & & & & \\
$T$   &  CN    & Spot &  Rel  & Strike &  Rel  & Hybrid &  Rel  & TE  &  Rel  & Hybrid &  Rel  & Hybrid &  Rel  \\
      &        &  VA  &  diff &  VA    &  diff &        &  diff &     &  diff &   VA   &  diff &  VA-2  &  diff \\
yrs & [$\$ $] & [$\$ $] & [$\%$] & [$\$ $] & [$\% $] & [$\$ $] & [$\%$] & [$\$ $] & [$\%$] & [$\$ $] & [$\%$] & [$\$ $] & [$\%$]  \\
& & & & & & & & & & & & & \\
\hline
& & & & & & & & & & & & & \\
1  & 2.18  & 3.14  & 43.9 & 2.18  & 0.2  & 2.18  & 0.1   & 5.86  & 168.6 & 2.19  & 0.2 & 2.18  & 0.2  \\
2  & 4.42  & 5.07  & 14.6 & 4.65  & 5.3  & 3.85  & -12.9 & 4.70  & 6.3   & 4.46  & 1.0 & 4.46  & 0.8  \\
3  & 6.76  & 7.12  & 5.9  & 7.40  & 10.1 & 5.90  & -12.2 & 5.69  & -15.3 & 6.81  & 1.2 & 6.77  & 0.6  \\
4  & 9.01  & 9.22  & 2.3  & 10.23 & 13.6 & 8.08  & -10.3 & 7.39  & -17.9 & 9.14  & 1.5 & 9.03  & 0.2  \\
5  & 11.23 & 11.30 & 0.6  & 13.05 & 16.2 & 10.28 & -8.5  & 9.35  & -16.7 & 11.41 & 1.6 & 11.22 & -0.1 \\
6  & 13.38 & 13.34 & -0.3 & 15.81 & 18.2 & 12.44 & -7.0  & 11.40 & -14.8 & 13.60 & 1.6 & 13.34 & -0.3 \\
7  & 15.45 & 15.33 & -0.8 & 18.50 & 19.8 & 14.55 & -5.8  & 13.45 & -13.0 & 15.70 & 1.6 & 15.38 & -0.5 \\
8  & 17.43 & 17.25 & -1.0 & 21.09 & 21.0 & 16.59 & -4.8  & 15.46 & -11.3 & 17.70 & 1.5 & 17.33 & -0.5 \\
9  & 19.32 & 19.10 & -1.2 & 23.58 & 22.0 & 18.54 & -4.0  & 17.40 & -9.9  & 19.61 & 1.5 & 19.20 & -0.6 \\
10 & 21.12 & 20.86 & -1.2 & 25.96 & 22.9 & 20.40 & -3.4  & 19.28 & -8.7  & 21.42 & 1.4 & 20.99 & -0.6 \\
11 & 22.84 & 22.56 & -1.2 & 28.25 & 23.7 & 22.19 & -2.9  & 21.08 & -7.7  & 23.15 & 1.4 & 22.69 & -0.7 \\
& & & & & & & & & & & & & \\
\hline
\end{tabular}}
\label{single_calls}
\end{table}
\begin{table}[h]
  \tbl{Comparison of different analytic methods with CN numerical results for a multi-dividend family of Calls. First column shows CN numerical results. In consequent columns different analytic approximations and their corresponding relative differences with respect to CN numerics are presented. These results corresponds to Fig. \ref{Fig2}.}
{\begin{tabular}{|c|c|c|c|c|c|c|c|c|c|c|c|c|c|} \hline
& & & & & & & & & & & & & \\
$T$   &  CN    & Spot &  Rel  & Strike &  Rel  & Hybrid &  Rel  & TE  &  Rel  & Hybrid &  Rel  & Hybrid &  Rel  \\
      &        &  VA  &  diff &  VA    &  diff &        &  diff &     &  diff &   VA   &  diff &  VA-2  &  diff \\
yrs & [$\$ $] & [$\$ $] & [$\%$] & [$\$ $] & [$\% $] & [$\$ $] & [$\%$] & [$\$ $] & [$\%$] & [$\$ $] & [$\%$] & [$\$ $] & [$\%$]  \\
& & & & & & & & & & & & & \\
\hline
& & & & & & & & & & & & & \\
1 & 10.19 &	10.20 &  0.0 & 10.23 & 0.3 & 10.18 & -0.1 & 10.19 &  0.0 & 10.20 & 0.1  & 10.20 & -0.0 \\
2 & 13.22 &	13.15 &	-0.5 & 13.33 & 0.8 & 13.16 & -0.4 & 13.21 & -0.1 & 13.22 & 0.1  & 13.21 & -0.1 \\
3 & 15.04 &	14.84 &	-1.4 & 15.29 & 1.7 & 14.91 & -0.8 & 15.02 & -0.2 & 15.05 & 0.0  & 15.02 & -0.1 \\
4 & 16.24 & 15.82 &	-2.6 & 16.67 & 2.8 & 16.01 & -1.4 & 16.21 & -0.2 & 16.24 & 0.0  & 16.22 & -0.2 \\
5 & 17.07 &	16.33 & -4.3 & 17.77 & 4.1 & 16.70 & -2.2 & 17.04 & -0.2 & 17.05 & -0.2 & 17.04 & -0.2 \\
6 & 17.66 &	16.51 &	-6.5 & 18.64 & 5.5 & 17.11 & -3.1 & 17.62 &	-0.2 & 17.60 & -0.4 & 17.62 & -0.2 \\
7 & 18.08 & 16.41 & -9.2 & 19.37 & 7.1 & 17.31 & -4.3 & 18.03 & -0.3 & 17.96 & -0.7 & 18.02 & -0.3 \\
8 & 18.37 & 16.06 & -12.5 &	19.99 & 8.8 & 17.34 & -5.6 & 18.32 & -0.3 & 18.17 & -1.1 & 18.31 & -0.3 \\
9 & 18.57 & 15.56 &	-16.2 & 20.53 &	10.5 & 17.25 & -7.1 & 18.52 & -0.3 & 18.27 & -1.7 & 18.50 & -0.4 \\
10 & 18.72 & 14.87 & -20.6 & 21.02 & 12.3 & 17.06 & -8.8 & 18.65 & -0.4 & 18.27 & -2.4 & 18.64 & -0.4 \\
11 & 18.81 & 14.01 & -25.5 & 21.47 & 14.1 & 16.79 & -10.7 & 18.73 & -0.4 & 18.21 & -3.2 & 18.73 & -0.4 \\
& & & & & & & & & & & & & \\
\hline
\end{tabular}}
\label{multi_calls}
\end{table}
\begin{table}[h]
  \tbl{Comparison of different analytic methods with CN numerical results for a multi-dividend family of Puts (Liquidator case).
  First column shows CN results. In consequent columns different analytic approximations and their corresponding relative differences with respect to CN numerics are presented. These results corresponds to Fig.\ref{Fig3}.}
{\begin{tabular}{|c|c|c|c|c|c|c|c|c|c|c|c|c|c|} \hline
& & & & & & & & & & & & & \\
$T$ & CN & Hybrid &  Rel  & Hybrid &  Rel  & Hybrid   &  Rel  &  Hybrid  &  Rel  & Hybrid &  Rel  & Hybrid &  Rel  \\
    &    &        &  diff &  PA    &  diff &    VA    &  diff &  VAPA  &  diff &   VA-2   &  diff &  VAPA-2  &  diff \\
yrs & [$\$ $]  & [$\$ $] & [$\%$] & [$\$ $] & [$\% $] & [$\$ $] & [$\%$] & [$\$ $] & [$\%$] & [$\$ $] & [$\%$] & [$\$ $] & [$\%$]  \\
& & & & & & & & & & & & & \\
\hline
& & & & & & & & & & & & & \\
1  & 13.10 & 13.09 & -0.1 & 13.09 &	-0.1 & 13.11 & 0.1  & 13.11 & 0.1 & 13.11 & 0.0 & 13.11 & 0.0 \\
2  & 18.86 & 18.81 & -0.3 & 18.81 & -0.3 & 18.87 & 0.1  & 18.87 & 0.1 & 18.86 & 0.0 & 18.86 & 0.0 \\
3  & 23.25 & 23.13 & -0.5 & 23.13 & -0.5 & 23.26 & 0.0  & 23.26 & 0.0 & 23.24 & -0.1 & 23.24 & -0.1 \\
4  & 26.87 & 26.66 & -0.9 & 26.64 & -0.9 & 26.89 & 0.0  & 26.87 & 0.0 & 26.87 & -0.1 & 26.86 & -0.1 \\
5  & 29.92 & 29.64 & -0.9 & 29.50 & -1.3 & 29.98 & 0.2  & 29.84 & -0.3 & 29.98 & 0.2 & 29.89 & -0.1 \\
6  & 32.33 & 32.20 & -0.4 & 31.69 & -2.0 & 32.69 & 1.1  & 32.13 & -0.6 & 32.71 & 1.1 & 32.31 & -0.1\\
7  & 34.06 & 34.42 & 1.0  & 33.21 & -2.5 & 35.07 & 2.9  & 33.74 & -1.0 & 35.13 & 3.1 & 34.06 & 0.0\\
8  & 35.12 & 36.36 & 3.5  & 34.10 & -2.9 & 37.19 & 5.9  & 34.70 & -1.2 & 37.33 & 6.3 & 35.16 & 0.1 \\
9  & 35.59 & 38.07 & 6.9  & 34.47 & -3.1 & 39.09 & 9.8 & 35.12 & -1.3 & 39.32 & 10.5 & 35.69 & 0.3 \\
10 & 35.56 & 39.58 & 11.3 & 34.42 & -3.2 & 40.79 & 14.7 & 35.09 & -1.3 & 41.15 & 15.7 & 35.73 & 0.5\\
11 & 35.14 & 40.90 & 16.4 & 34.02 & -3.2 & 42.32 & 20.4 & 34.71 & -1.2 & 42.84 & 21.9 & 35.38 & 0.7\\

& & & & & & & & & & & & & \\
\hline
\end{tabular}}
\label{multi_puts}
\end{table}

Table \ref{single_calls} presents the results for single dividend family of Calls. Although different approximations typically perform well in certain limits and a lot worse in other limits, the newly developed Hybrid VA and Hybrid VA-2 approximations work consistently well for all terms $T$. Surprisingly TE approach is the worst performer for this example producing unsatisfactory results in smaller $T$ limit. Perhaps the size of disagreement can be reduced if more perturbation orders of TE approach are taken into account (we had a code supplied by the authors of \cite{VeigaW2009} where only the first two orders were available), but these tests are beyond the scope of the current work. Here we don't present the results for Puts, because for this particular single dividend example the parity violation amount is rather small ($\Delta P_L \approx 0.04$) and the quality of Put approximations is approximately defined by the corresponding Call results via the standard parity relation (\ref{parity_BS}) which approximately holds. Note, however, that for a single dividend case parity violation value may be a lot higher for Survivor dividend policy case. For example, for the single dividend family considered in Table \ref{single_calls} the corresponding Survivor parity violation value $\Delta P_S \approx 0.42 \gg 0.04$.

Table \ref{multi_calls} presents the results for multi-dividend family of Calls. The newly developed Hybrid VA and Hybrid VA-2 approximations work consistently well for all terms $T$, although the simpler Hybrid VA method starts to diverge slightly from numerics at $T > 8$. For this example TE approach works as well as Hybrid VA-2 (both methods are producing essentially identical results).

Table \ref{multi_puts} presents the results for multi-dividend family of Puts for CN numerics and a few better performing analytic approaches for Liquidator dividend policy case. Note that for parity violation adjustment (PA), $\Delta P$ calculations for Hybrid PA results are based on Hybrid method Put formulas; $\Delta P$ calculations for Hybrid VAPA results are based on Hybrid VA method Put formulas; and  $\Delta P$ calculations for Hybrid VAPA-2 results are based on Hybrid VA-2 method Put formulas. Clearly PA methods substantially outperform their non-PA analogs. Also, similarly to the corresponding multi-dividend Call results, Hybrid VAPA and Hybrid VAPA-2 are clear leaders in approximation accuracy.
\section{Discussion}

In conclusion, we reviewed existing analytic approximations for vanilla European options with discrete dividends and suggested a novel algorithm, which typically provides a superior accuracy in comparison to other approaches. In addition, we spent considerable efforts in clarifying the correct choice of boundary conditions for a benchmark numerical algorithm: a finite difference Crank-Nicolson scheme. Finally we comprehensively investigated a rarely mentioned Put-Call parity violation phenomenon, and successfully applied our newly developed analytical approach for calculation of the corresponding parity violation adjustment (needed for any BS-style approximation for Puts with discrete dividends). We have not reported the related results for Greeks, but the simple form of our approximation allows for obtaining them in a rather straightforward fashion.

We note, that in the simpler representation our method generalises heuristic volatility correction formulas of \cite{BenederW2001}] and does not allow calculation of the next perturbation order to gain extra accuracy. However, we have also presented a different version of our approach which utilises advantages of the more involved method of \cite{BosGS2003} which, in principle, may allow for calculation of higher perturbation orders (see Appendix A for details).

Also we believe that the presented ideas may be useful for developing better fast algorithms/approximations for other types of options with discrete dividends, including barrier options, American options, etc. Call results can be directly used for calculation of local volatility surfaces [as in, e.g., \cite{Wilmott2006}].

\section*{Acknowledgements}

Authors are indebted to NAB Wholesale Banking MRQS group for support and useful suggestions and especially to Volf Frishling and Stephen Edney for pointing out the general topic of this work and many insightful discussions. Authors are also grateful to Uwe Wystup and Carlos Veiga for sharing their TE approach code.

\section*{Appendix A} \label{AppendixA}

Derivation details of \cite{BosGS2003} style Hybrid VA-2 formulas:

Consider the generalised hybrid framework:
\[
0\leq \alpha_i\leq 1,\ {\hat{D}_S(t) \equiv \sum_{t < t_i \le T} \alpha_i d_i \exp{(-r (t_i-t))}}\ \text{and}\ {\hat{D}_K(t) \equiv \sum_{0 \le t_i \le t} (1-\alpha_i) d_i \exp{(-r (t_i-t))}}
\]
and the adjusted stock process
\[
\hat{S}_t=S_t-\hat{D}_S(t)+\hat{D}_K(t),\ d\hat{S}_t=r\hat{S}_tdt+\hat{\sigma}(\hat{S}_t,t)\hat{S}_tdW_t.
\]
In particular, $\alpha_i=1$ gives the spot approximation model, $\alpha_i=0$ gives the strike approximation model, and $\alpha_i=(T-t_i)/T$ gives the hybrid approximation model.

The local volatility can then be written in terms of the constant volatility
\[
\hat{\sigma}(\hat{S}_t,t)=\sigma\left(1+\frac{\hat{D}_S(t)-\hat{D}_K(t)}{\hat{S}_t}\right).
\]
By perturbation theory (see Appendix B in \cite{BosGS2003}), the implied volatility $\sigma(K,T)$ can be approximated by
\[
\sigma(K,T)^2=\frac{1}{T}\int_0^T E\hat{\sigma}\left(\exp\left(rt+X_{\sigma^2 t}^{s, l ,k}\right),t\right)^2 dt.
\]
where $X_{\sigma^2 t}^{s, l ,k}$ is the time $\sigma^2 t$ value of a Brownian bridge from $s=\ln \left(S_0-\hat{D}_S(0)\right)$ to $k=\ln \left(\left(K+\hat{D}_K(t)\right)\exp(-rT)\right)$ of length $l=\sigma^2 T$. Using the properties of the brownian bridge
\[
rt+X_{\sigma^2 t}^{s, l ,k}\sim \mathcal{N}\left(x_t, \sigma^2 y_t\right)\ \text{with}\ x_t=\left(s+\frac{(k-s)t)}{T}\right)+rt,\ y_t=\frac{t(T-t)}{T},
\]
expand the expectation
\begin{align*}
&E\hat{\sigma}\left(\exp\left(rt+X_{\sigma^2 t}^{s, l ,k}\right),t\right)^2\\
={}& \sigma^2\left(1+2\left(\hat{D}_S(t)-\hat{D}_K(t)\right)E\left(e^{-\left(rt+X_{\sigma t}^{s, l ,k}\right)}\right)+\left(\hat{D}_S(t)-\hat{D}_K(t)\right)^2E\left(e^{-2\left(rt+X_{\sigma t}^{s, l ,k}\right)}\right)\right)\\
={}& \sigma^2\left(1+2\left(\hat{D}_S(t)-\hat{D}_K(t)\right)e^{-x_t+\frac{\sigma^2}{2}y_t}+\left(\hat{D}_S(t)-\hat{D}_K(t)\right)^2e^{2\left(-x_t+\sigma^2y_t\right)}\right).
\end{align*}
Finally, by using the following identities
\begin{align*}
\int_0^{\tau}e^{rt} e^{-x_t+\frac{\sigma^2}{2}y_t} dt &= e^{\frac{a^2}{2}-s}\int_0^{\tau} e^{-\frac{\left(\frac{t\sigma}{\sqrt{T}}-a\right)^2}{2}} dt=\frac{\sqrt{2\pi T}}{\sigma}e^{\frac{a^2}{2}-s}\left(\Phi(a)-\Phi\left(a-\frac{\sigma \tau}{\sqrt{T}}\right)\right)\\
\int_0^{\tau}e^{2rt} e^{2\left(-x_t+\sigma^2y_t\right)} dt &= e^{\frac{b^2}{2}-2s}\int_0^{\tau} e^{-\frac{\left(\frac{2t\sigma}{\sqrt{T}}-b\right)^2}{2}} dt=\frac{1}{\sigma}\sqrt{\frac{\pi T}{2}}e^{\frac{b^2}{2}-2s}\left(\Phi(b)-\Phi\left(b-\frac{2\sigma \tau}{\sqrt{T}}\right)\right),
\end{align*}
where
\[
a=\frac{s-k}{\sqrt{T}\sigma}+\frac{\sqrt{T}\sigma}{2},\ b=\frac{s-k}{\sqrt{T}\sigma}+\sqrt{T}\sigma
\]
and $\Phi$ is the cumulative Gaussian distribution function, the closed form of the implied volatility $\sigma(K,T)^2$ can be computed to be
\begin{align*}
\sigma(K,T)^2={}&\frac{\sigma^2}{T}\int_0^T \Bigg(1+2\left(\hat{D}_S(t)-\hat{D}_K(t)\right)e^{-x_t+\frac{\sigma^2}{2}y_t}+\left(\hat{D}_S(t)-\hat{D}_K(t)\right)^2e^{2\left(-x_t+\sigma^2y_t\right)}\Bigg) dt\\
={}&\sigma^2+2\sigma\sqrt{\frac{2\pi}{T}}e^{\frac{a^2}{2}-s}\Bigg[\sum_i \alpha_i d_i e^{-rt_i}\left(\Phi(a)-\Phi\left(a-\frac{\sigma t_i}{\sqrt{T}}\right)\right)\\
&-\sum_i (1-\alpha_i) d_i e^{-rt_i}\left(\Phi\left(a-\frac{\sigma t_i}{\sqrt{T}}\right)-\Phi\left(a-\sigma\sqrt{T}\right)\right)\Bigg]\\
&+\sigma\sqrt{\frac{\pi}{2T}} e^{\frac{b^2}{2}-2s}\Bigg[\sum_{i,j}\alpha_i d_i \alpha_j d_j e^{-r(t_i+t_j)}\left(\Phi(b)-\Phi\left(b-\frac{2\sigma\min(t_i, t_j)}{\sqrt{T}}\right)\right)\\
&+\sum_{i,j}(1-\alpha_i) d_i (1-\alpha_j) d_j e^{-r(t_i+t_j)}\left(\Phi\left(b-\frac{2\sigma\max(t_i, t_j)}{\sqrt{T}}\right)-\Phi\left(b-2\sigma\sqrt{T}\right)\right)\\
&-2\sum_{i>j}\alpha_i d_i (1-\alpha_j) d_j e^{-r(t_i+t_j)}\left(\Phi\left(b-\frac{2\sigma t_j}{\sqrt{T}}\right)-\Phi\left(b-\frac{2\sigma t_i}{\sqrt{T}}\right)\right)\Bigg].
\end{align*}

\markboth{Alexander Buryak and Ivan Guo}{Applied Mathematical Finance}
%\medskip
%\noindent This list was produced by:
%\medskip

\label{lastpage}

\end{document}